\documentclass[conference]{IEEEtran}
\IEEEoverridecommandlockouts
\usepackage[nocompress]{cite}
\usepackage{amsmath,amssymb,amsfonts}
\usepackage{algorithmic}
\usepackage{graphicx}
\usepackage{textcomp}
\usepackage{xcolor}
\usepackage{nameref}

\usepackage{url}
\usepackage{multirow}
\def\BibTeX{{\rm B\kern-.05em{\sc i\kern-.025em b}\kern-.08em
    T\kern-.1667em\lower.7ex\hbox{E}\kern-.125emX}}
\begin{document}

\title{RVCoreP-32IM: An effective architecture to implement mul/div instructions for ﬁve stage RISC-V  soft processors}

\author{\IEEEauthorblockN{Md Ashraful Islam}
\IEEEauthorblockA{\textit{ School of Computing,} \\
\textit{Tokyo Institute of Technology}\\
ashraful@arch.cs.titech.ac.jp 
}
\and
\IEEEauthorblockN{Hiromu MIYAZAKI}
\IEEEauthorblockA{\textit{ School of Computing,} \\
\textit{Tokyo Institute of Technology}\\
miyazaki@arch.cs.titech.ac.jp 
}
\and
\IEEEauthorblockN{Kenji KISE}
\IEEEauthorblockA{\textit{ School of Computing,} \\
\textit{Tokyo Institute of Technology}\\
kise@c.titech.ac.jp}
}
\maketitle

\begin{abstract}

RISC-V, an open instruction set architecture, is getting the attention of soft processor developers. Implementing only a basic 32-bit integer instruction set of RISC-V, which is defined as RV32I, might be satisfactory for embedded systems. However, multiplication and division instructions are not present in RV32I, rather than defined as M-extension. Several research projects have proposed both RV32I and RV32IM processor. However, there is no indication of how much performance can be improved by adding M-extension to RV32I. In other words, when we should consider adding M-extension into the soft processor and how much hardware resource requirements will increase.

In this paper, we propose an extension of the RVCoreP soft processor (which implements RV32I instruction set only) to support RISC-V M-extension instructions. A simple fork-join method is used to expand the execution capability to support M-extension instructions as well as a possible future enhancement. We then perform the benchmark using Dhrystone, Coremark, and Embench programs. We found that RV32IM is 1.87 and 3.13 times better in performance for radix-4 and DSP multiplier, respectively. In addition to that, our RV32IM implementation is 13\% better than the equivalent RISC-V processor. 

\end{abstract}

\begin{IEEEkeywords}
Microarchitecture, Soft processor, RISC-V Embedded systems, Multiplication/Division instruction
\end{IEEEkeywords}

\section{Introduction}

Many embedded systems are built using FPGA for research and product development, which increases the number of soft processor uses. While hard processors are available in modern FPGA, which are much faster than the soft processors and do not consume LUT, but they have a few drawbacks \cite{b12}. Firstly, the number of built-in hard processors in the FPGA may be sub-optimal or redundant for the application. Secondly, the configuration of each processor is fixed and is not configurable according to the application. Thirdly, each hard processor is placed in a fixed position, which may lead to difficulties to route between the processors and the custom logic or accelerators. 

On the other hand, the soft processor cannot easily match the performance/area/power of a hard processor, but soft processors have several advantages. Designers can implement the required number of soft processors for their application, and EDA tools can be used with different options to find the optimal place and routing to maximize the throughput. Moreover, the developers can choose the implementation of each soft processor with different configurations as well as tightly integrate the custom logic or acceleration functions to compete with the complexity of the application.

There are researches on soft processor with the configurable instruction set. However, a soft processor with a fully custom instruction set is difficult to use; as the software toolchain must build for that from scratch. On the other hand, RISC-V is an open standard instruction set with several extensions for 32-bit, 64-bit, and 128-bit. Among them, RV32I is basic (minimum integer) instructions that must have to be supported for compliance with RISC-V. In this research, we targeted for 32-bit soft processor based on RISC-V instruction set architecture (ISA). From the integer computational perspective, the next ISA extension of RISC-V to consider after RV32I should be M-extension, which relies on hardware multiplication and division, whereas RV32I performs the multiplication and division as a software routine. 

Multiplication operations are significant for many applications in the domain of digital signal processing, image processing, scientific computing, and many more. Many FPGA vendors provide embedded hard multiplier blocks as DSP blocks (which can perform a multiply-accumulate function) into the fabric. However, the number of such multiplier/DSP blocks is limited compared to the LUT count, and their operands bit length is fixed. In terms of resource usage, sequential multipliers could provide smaller LUT utilization than the tree-based multipliers. In this study, we have implemented the hardware multiplication unit for RV32M multiplication instructions in both ways (i.e., utilizing DSP block and utilizing sequential logic element only). 

For some scientific computing and special purpose operations like graphics rendering, etc., division operations are more frequent. For hardware division, digit recurrence division algorithms are the simplest and most widely used for hardware implementation. They usually require less number of LUT but have higher latency. In this study, we used the digit recurrent division algorithm for simplicity. 

In this work, we present an efficient way to integrate hardware multiplier and divider to support M-extension instructions on a 5-stage in-order soft processor. We have named our proposed soft processor microarchitecture as RVCoreP-32IM. RVCoreP-32IM is written in Verilog HDL, which is configurable in the context of supported instructions from RV32I to RV32IM. We then implemented it on Artix-7 FPGA and measured the LUT count and performance. The rest of this paper is organized as follows - section \ref{background} provides the related research works, section \ref{proposal} describes the proposed microarchitecture details and section \ref{evaluation} presents the experiment results. 

\section{Background} \label{background}

Commercial soft processors, like the NIOSII \cite{nios}, MicroBlaze \cite{mblaze}, and ARM Cortex-M1\cite{m1}, are optimized for FPGA device-specific resource usage and frequency, but they are not open source. As a result, architectural exploration from different perspectives with the soft processors is difficult. In contrast to commercial ISA, the RISC-V\cite{riscv} is an open-source ISA with available open-source software tools, libraries, and applications from open source community. 

Several kinds of research have been conducted for RISC-V processor architecture development. The Berkeley Rocket chip generator \cite{rocket} makes different instances of the Rocket RISC-V core. However, that is written in Chisel language, which is different from widely used Verilog or VHDL language. Similarly, BOOM \cite{boom}, which is an out-of-order processor from Berkeley, is also written in Chisel. Developers may not be accustomed to functional verification of Chisel compared to SystemVerilog, UVM, or C/C++ based verification. Apart from this, PULP project \cite{pulp} has developed RISC-V cores RI5CY \cite{ri5cy} and Ariane \cite{ariane} in SystemVerilog for low power SoC. However, those cores are not targeted for FPGA in terms of soft processor. FPGA is significantly different from ASIC; for example, logic is implemented in fixed-size LUT, on-chip memories are arranged as fixed-width blocks, hard multiplication or DSP blocks of fixed size, etc. 

The research works on RISC-V based FPGA soft processor is also growing like GRVI \cite{grvi}, PicoRV32 \cite{picorv}, Taiga \cite{taiga} VexRiscv \cite{vex} and RVCoreP \cite{rvcorep}. The GRVI is an FPGA-based accelerator framework based on scalar RISC-V cores of a 2/3-stage pipeline supporting only the RV32I instruction set and that coupled with accelerators. Similarly, PicoRV32 is RV32I with an optional M and C extension focused on the smaller logic resources but not optimized for the performance. Taiga is a RISC-V soft processor consisting of variable latency parallel execution units, which has higher performance but with additional logic overhead. VexRiscv, as a soft processor, has a good balance of performance and logic resource utilization. Again VexRiscv is written in Scala. RVCoreP, which is written in Verilog, is found to have better performance than VexRiscv for RV32I. However, RVCoreP does not support M extension. 

In this paper, we have extended the RVCoreP for supporting M extension instructions. We also studied the performance gain of using M extension over RV32I base instructions. Our motivation for this study is to find the optimum implementation of M-extension instruction and the impact of hardware multiplication and division by using Dhrystone\cite{dhrystone}, CoreMark\cite{coremark}, and Embench\cite{embench} benchmark programs.

\section{Proposal} \label{proposal}

\begin{figure}[tb]

\centering

\includegraphics[width=0.8\linewidth]{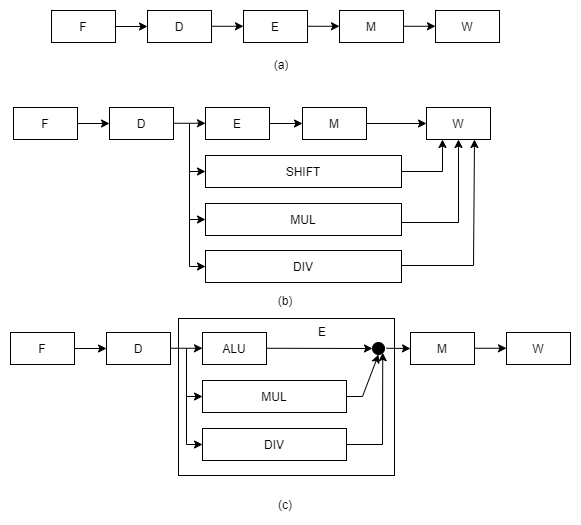}

\caption{The five stage Pipelining configurations where a) RV32I, b) RV32IM on VexRiscv, and c) RV32IM on RVCoreP-32IM}

\label{fig1}

\end{figure}

For this study, we have chosen the in-order scalar processor as out-of-order processors are complex and consume significant FPGA resources. Figure \ref{fig1} illustrates the 5-stage pipeline diagram, where F for fetch, D for decode, E for execute, M for memory, MUL for a multiplier, DIV for a divider, SHIFT for the shifter, and W for the write back. For the RV32I implementation, as shown in figure \ref{fig1}(a), it is easy to consider that every stage will take only one clock cycle (considering Load/Store instruction will be executed in a single cycle with Block RAM). However, for M-extension, where multiplication and division should be performed in hardware, is not reasonable to consider as a single clock cycle unit from the maximum operating frequency perspective. 

\begin{figure}[tb]

\centering

\includegraphics[width=0.8\linewidth]{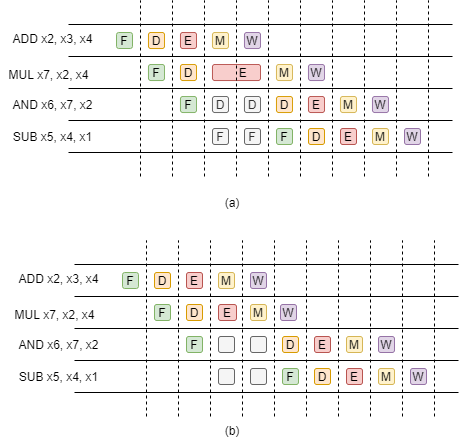}

\caption{Pipeline execution flow of a) RVCoreP-32IM, b) VexRiscv}

\label{fig1_2}

\end{figure}

We took VexRiscv as a reference processor, which has a shifter, multiplier, and divider as a separate functional unit other than as part execution unit. Figure \ref{fig1}(b) shows that the results of shifter, multiplier, and divider are committed to the write back stage. In our proposed RVCoreP-32IM, we consider the execution stage as a multi clock cycle unit. During the multiplication and division operation, the execution stage will consume multiple cycles, but for the other integer instructions, it should take only a single clock cycle. Figure \ref{fig1}(c) represents this idea, as there is a fork in the execution unit to accommodate integer ALU, multiplication, and division unit and then joined into the memory stage. When multi-cycle execution occurs, the decode and fetch pipeline will stall until the instruction moves to the memory stage. 

In VexRiscv, shift, multiplication and division results are available in the write back stage. When there is no data dependency in the instruction sequence, then shift and multiply instruction is pipelined without bubble, and divide or reminder instruction takes 32 cycles. But if the next instruction after a shift or multiply or divide/reminder has data dependency on the previous instruction result, then the two-cycle bubble is inserted in the pipeline. On the other hand, in our proposal, RVCoreP-32IM shift always executes in a single cycle. Multiplication instruction stalls fetch and decode stage 1 cycle for DSP based multiplier and 17 cycles for the radix-4 multiplier. Divide/reminder instruction stalls fetch and decode stage for 34 cycles. For both multiplication and divide/reminder instruction, if there is data-dependent instruction immediately after that instruction, then the fetch and decode stage stalls one more cycle. 

As an example, we have shown the execution of multiplication instruction on RVCoreP-32IM in figure \ref{fig1_2}(a). In this example, multiplication takes two cycles in the execution stage, and the next instruction uses the result of multiplication instruction (in this example, AND instruction uses register x7). As a result, subsequent instructions after multiplication (AND \& SUB instruction in the example) stalled in the Fetch and Decode stage for two cycles, respectively. As a reference, we have shown the VexRiscv execution in \ref{fig1_2}(b). When there is a data-dependent instruction after the load instruction, like VexRiscv, RVCoreP-32IM inserts two-cycle bubbles.

\subsection{Microarchitecture of RVCoreP-32IM}\label{microarch}

The soft processor microarchitecture is based on RVCoreP. Our proposed microarchitecture combines the multiplier and divider into the execution unit data path. When M-extension is disabled from the configuration, then this architecture resembles RVCoreP.

\begin{figure*}[tb]

\centering

\includegraphics[clip,width=0.8\linewidth]{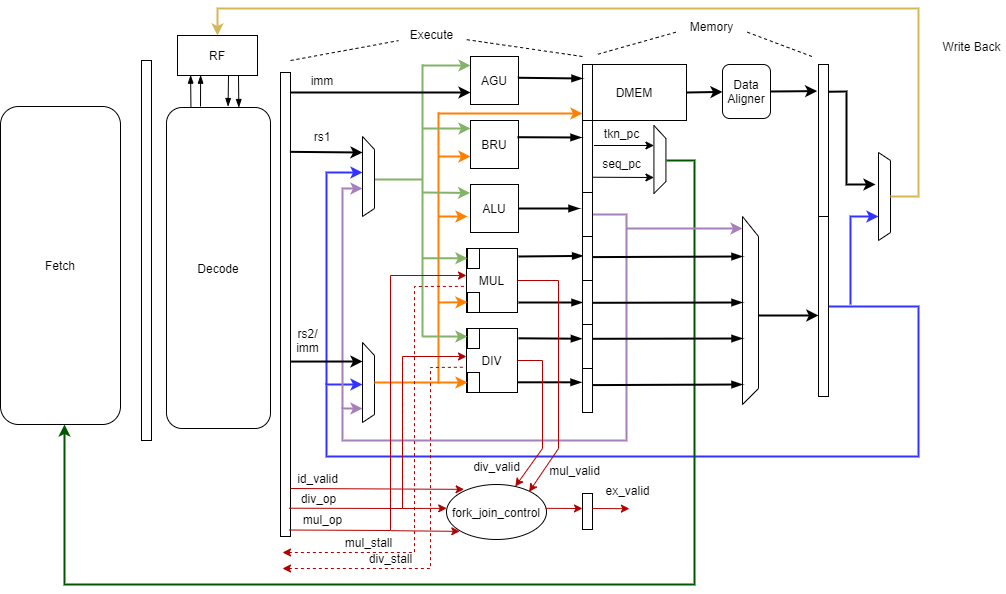}

\caption{Microarchitecture of RVCoreP-32IM}

\label{prconame}

\end{figure*}

The proposed microarchitecture of RVCoreP-32IM is shown in Figure \ref{prconame}. AGU is an address generation unit for memory instructions, and BRU is a branch unit for branch instructions, and ALU is the arithmetic and logical unit for RV32I integer arithmetic and logical instructions, MUL is a multiplier for multiplication instructions, and DIV is a divider for division and reminder instruction execution.

The instruction fetch unit uses the gshare branch prediction, which is configurable to be in pipeline fashion as proposed in RVCoreP. According to RVCoreP, the fetch and decode stage is stalled only if there is a data dependency hazard due to load instructions. Since the proposed architecture has a multi-cycle unit (multiplier and divider) and as this is in order scalar processor, so the fetch and decode unit must have to be stalled when M extension instructions are being executed. So the execution stage contains single-cycle ALU and multi-cycle multiplication and division unit. To facilitate this kind of arrangement, we use a simple fork and join, which is based on “valid” and “stall” signals. 

The “valid” signal propagates forward to indicate that new data is available for the next stage. And the “stall” signal propagates backward to signal the previous stage to halt, alternatively, we can say to insert a bubble. In RVCoreP-32IM, the decode stage asserts the “id\_valid” signal when a valid instruction is decoded. Now, if the decoded instruction is RV32I type instruction, then in the next clock cycle “ex\_valid” signal will be asserted to indicate valid data for memory stage. But if RV32M extensions instruction is decoded by the decode stage, then the “mul\_op” or “div\_op” signal will be asserted from the decode stage for multiplication and division instruction, respectively. When the “mul\_op” or “div\_op” signal will be asserted the rs1 and rs2 register’s value (after data forwarding multiplexer) will be captured by the multiplication or division unit in the next clock cycle. And “ex\_valid” will not be asserted in the next clock cycle of “id\_valid” if “mul\_op” or “div\_op” is asserted because the execute stage will need multiple clock cycle to finish this execution. When “mul\_op” or “div\_op” is asserted, then “mul\_stall” and “div\_stall” signals are asserted respectively and feedback to the decode stage. These stall signals remain asserted until the multiplication or division operation finishes. When any stall signal is asserted, the decoder stage stalls the fetch stage and inserts a bubble into the pipeline register between the decode and execute stage. 

For FPGA soft processor, the data forwarding (or feedback) path becomes the critical path for the timing \cite{b5}. To increase the operating frequency of the RVCoreP-32IM, multiplier, and divider output in the execution stage is not included in the data forwarding path from the memory stage to the execute stage. Two consecutive data-dependent instructions can be forwarded from the memory stage if the precedent instruction is ALU instruction. When there is data-dependency on a multiplication or division/reminder instruction, the fetch and decode stage will be stalled for one more cycle, so that data can be forwarded from the write back stage. In a typical five-stage processor, if there is a data dependency on load instruction, then one cycle bubble is inserted so that data can be forwarded from the write back stage. In our architecture, we did not include the load data into the forwarding path from the write back stage. As a result, when there is a data-dependent instruction after load instruction, two cycles bubble are inserted into the execute stage.

\subsection{Multiplier Unit}\label{multiply_s}

\begin{figure}[tb]

\centering

\includegraphics[width=0.8\linewidth]{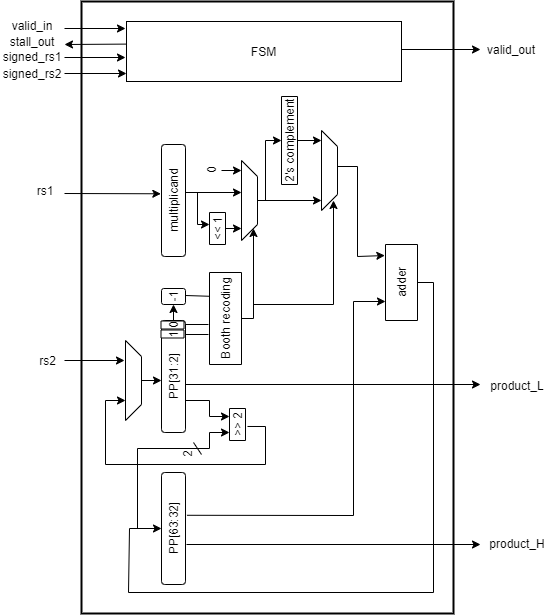}

\caption{Block diagram of radix-4 Booth multiplier}

\label{fig:mul_booth}

\end{figure}

\begin{figure}[tb]

\centering

\includegraphics[width=0.8\linewidth]{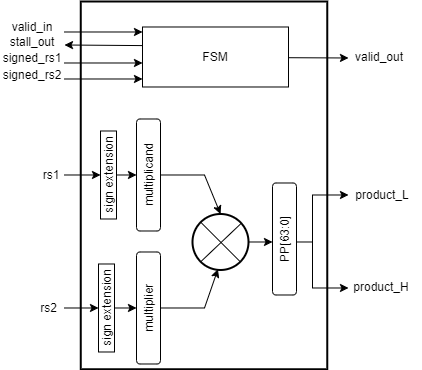}

\caption{Block diagram of DSP multiplier}

\label{fig:mul_dsp}

\end{figure}

RVCoreP-32IM has an option to choose between iterative radix-4 booth multiplier and DSP based multiplier and thus instantiated into the execute stage. The input and output of the multiplier unit are the same regardless of the implementation option, which is useful to experiment with other types of multipliers in the future. Figure \ref{fig:mul_booth} shows the radix-4 booth multiplier and figure \ref{fig:mul_dsp} shows the DSP multiplier diagram. The multiplier unit’s “valid\_in” signal is connected to “mul\_op” and “stall\_out” is connected to the “mul\_stall” of the execution block. When the multiplication is done “valid\_out” signal is asserted, and “stall\_out” is de-asserted. RISC-V multiplication instruction can perform both signed and unsigned multiplication and can use either a higher 32-bit or lower 32-bit product. The multiplication unit contains the data path and control path, which is a finite state machine(FSM) to control the data path. 

For the radix-4 booth multiplier, the rs1 register value is loaded into the multiplicand register, and the rs2 register value is loaded into the partial product (PP) register. As this is an iterative radix-4 multiplier, so the 2-bit product is computed in every clock cycle, and the PP register is 2-bit right-shifted in each clock (until multiplication is done), and depending on the sign values, one cycle is required for sign correction. This means the multiplication operation latency becomes 17 cycles (i.e. “stall\_out” will be asserted for 17 cycles), after that, “valid\_out” will be asserted. 


On the other hand, for the DSP multiplier, we used the DSP block, which is available on the FPGA. In the FPGA synthesis tool, when multiplication operand is used in RTL, it usually infers the FPGA DSP blocks to implement the multiplication operation. Although the Xilinx DSP block can execute multiplication in a single cycle, we used the pipeline register to the input of the DSP. Using this pipeline register at the input of DSP shortens the critical forwarding paths from the memory and the write Back stage to DSP. In our case, we used the multiplication operand in RTL as a datapath element and simple FSM to handle the input and output from the data path. Register values of rs1 and rs2 are loaded into multiplicand and multiplier register at the positive edge of the clock, respectively. Then DSP operation is done in a single cycle and stored into the PP register. 


As the DSP multiplier uses one clock cycle to capture into multiplicand and multiplier register and one cycle for computation, so the multiplication latency becomes 2 clock cycle. For the DSP multiplier, the “mul\_stall” signal is asserted for one clock. To summarize, multiplication operation latency for the execution stage is 18 and 2 clock cycles for radix-4 and DSP multiplier, respectively. 

\subsection{Divider Unit}\label{div_s}

\begin{figure}[tb]

\centering

\includegraphics[width=0.8\linewidth]{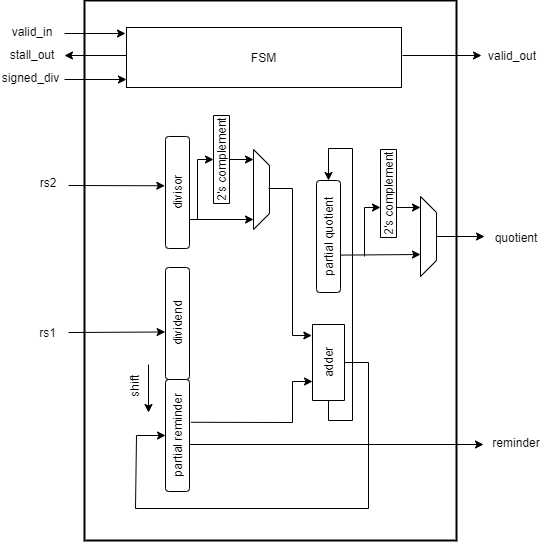}

\caption{Block diagram of radix-2 non-restoring divider}

\label{fig:div}

\end{figure}

The divider is implemented using an iterative non-restoring division algorithm. According to \cite{fpga:divider:bib}, we found that iterative non-restoring divider consumes the lowest logical resource and highest clock frequency after the SRT radix-2 divider. In this study, we used the smallest LUT utilization option, which is an iterative non-restoring divider. As shown in figure \ref{fig:div}, the divider unit has FSM for controlling the data path similar to the multiplier unit. The divider unit’s “valid\_in” and “stall\_out” are connected to “div\_op” and “div\_stall” respectively and “valid\_out” is connected to “div\_valid” in the execute stage. RISC-V division instruction has both signed and unsigned division and can use either quotient or reminder. If the dividend or divisor is zero, then the divider stalls for two clock cycles; otherwise, it stalls the 33 or 34 clocks depending on the sign, and then “valid\_out” will be asserted. Register values of rs1 and rs2 are loaded into the dividend and divisor register, respectively.

\section{Evaluation} \label{evaluation}

RVCoreP-32IM processor is written in Verilog HDL. A configuration option is provided as a header file. We have evaluated four processor configurations - RV32I, RV32IM(radix-4) where the multiplier is implemented as radix-4 booth multiplier, and RV32IM(DSP) where the multiplier used FPGA DSP blocks and VexRiscv with DSP based multiplier. We used Dhrystone \cite{dhrystone}, Coremark \cite{coremark} and Embench \cite{embench} benchmark programs for simulation. The evaluation focuses on two main aspects: (1) FPGA resource utilization and achieved maximum frequencies (2) how much performance gain is achieved for RV32IM by comparing with RV32I and VexRiscv while taking into account of achieved frequency. We calculated the performance gain by comparing the required time for execution of the same benchmark program by RV32I with other configurations. We then compared our result with VexRiscv in terms of execution time. Since VexRiscv has many possible configurations, so we took the maximum performance configuration for RV32IM without MMU and cache, and enabled dynamic branch prediction, DSP multiplier, and all bypassing options for register.

\subsection{FPGA implementation}

\begin{figure}[tb]

\centering

\includegraphics[width=0.8\linewidth]{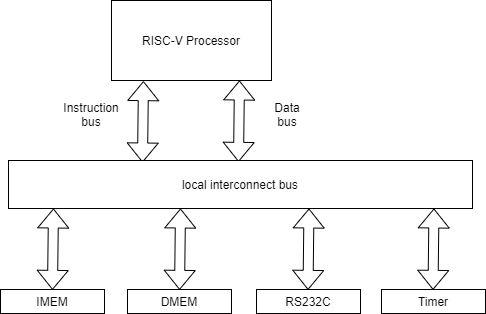}

\caption{Simple SoC for evaluation in FPGA}

\label{fig:soc}

\end{figure}

We implement three configurations of the proposed processor and one configuration of VexRiscv, then find the maximum operating frequency and hardware resource utilization.  RVCoreP-32IM used two-stage pipeline gshare branch prediction with a pattern history table (PHT) of 8,192 entries and a BTB of 512 entries. We prepare a simple SoC to evaluate the FPGA implementation of these processors. SoC block diagram is shown in figure \ref{fig:soc}, where instruction memory (IMEM), data memory (DMEM), timer, and RS-232C serial communication device are connected through a simple local interconnect bus. RISC-V program is loaded into IMEM and the processor starts execution from there. 

\begin{table*}

\centering

\caption{RVCoreP-32IM FPGA (Artix-7 speed grade -1) implementation result}

\label{fpga:table}

\centering


\begin{tabular}{|c|c|c|c|c|c|c|c|c|c|}

\hline

\multirow{2}{*} {Configuration} &

\multicolumn{2}{c|}{Frequency(MHz)} & 

\multicolumn{2}{c|}{Slice LUT} & 

\multicolumn{2}{c|}{Slice register} & 

\multicolumn{2}{c|}{Slice} & 

\multirow{2}{*} {DSP} \\ 


\cline{2-9} 


& 4KB & 64KB & 4KB & 64KB & 4KB & 64KB & 4KB & 64KB & \\ 

\hline

   RV32I           & 174 & 164 & 958  & 958 & 647 & 641 & 391 & 396 & 0 \\ 

\hline

   RV32IM(radix-4) & 169 & 162 & 1456 & 1457 & 970 & 970 & 548 & 565 & 0 \\ 

\hline

   RV32IM(DSP)     & 169 & 162 & 1331 & 1338 & 901 & 903 & 487 & 539 & 4 \\ 

\hline

  VexRiscv(RV32IM) & 153 & 143 & 1367 & 1394 & 1044 & 1013 & 533   & 503 & 4 \\ 

\hline

\end{tabular}

\end{table*}

We evaluated the operating frequency and hardware resource utilization for Nexys A7 board \cite{nexysa7} containing xc7a-100tcsg324-1 FPGA, which is a family of Xilinx Artix-7 FPGA of speed grade -1. Xilinx Vivado 2018.3 is used to evaluate the operating frequency and hardware resource utilization.

Flow\_PerfOptimaized\_high strategy is used for logic synthesis, and Performance\_ExplorePostRoutePhysOpt strategy is used for placement and routing. We performed the logic synthesis and placement and routing by incrementally stepping the clock cycle constraint in the 1MHz scale. The constraints for the highest frequency that achieves the positive timing slack are used as the operating frequency in the table. For hardware resource evaluation, we used the result of placement and routing at the maximum operating frequency. We prepared the FPGA implementation result for both IMEM and DMEM size to 4KB and 64KB. Since Xilinx FPGA BRAM size is 36Kb, we choose 4KB as a lower memory size. And to satisfy the memory requirement to run all the benchmark programs we took 64KB as the upper memory size for the evaluation.

\begin{figure}[t]

\centering

\includegraphics[width=0.6\linewidth]{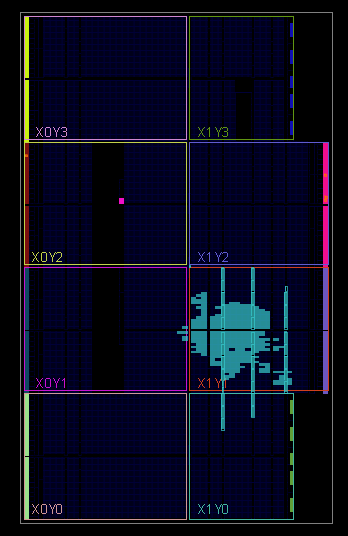}

\caption{RV32I implementation for 64KB}

\label{fig:rv32i_pnr}

\end{figure}

\begin{figure}[tb]

\centering

\includegraphics[width=0.6\linewidth]{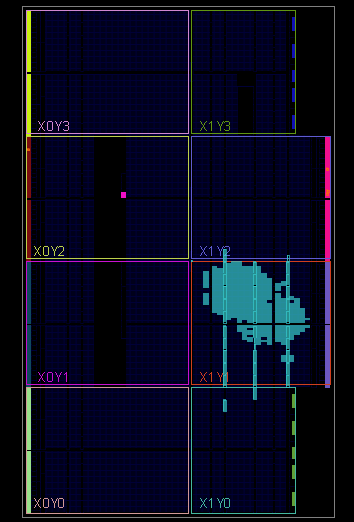}

\caption{RV32IM(radix-4) implementation for 64KB}

\label{fig:rv32im1_pnr}

\end{figure}

\begin{figure}[tb]

\centering

\includegraphics[width=0.6\linewidth]{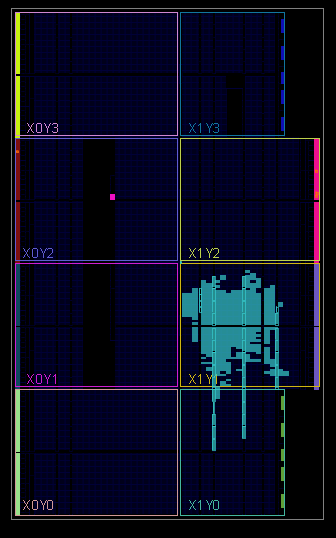}

\caption{RV32IM(DSP) implementation for 64KB}

\label{fig:rv32im3_pnr}

\end{figure}

Implementation results are shown in table \ref{fpga:table}. From that table, we can see that LUT resource utilization increases around 52\% for RV32IM(radix-4) and around 40\% for RV32IM(DSP) compared to RV32I implementation. On top of that, RV32IM(DSP) uses 4 DSP blocks of FPGA. RV32I achieves a bit higher frequency than RV32IM (regardless of multiplier type) for 4KB memory size. But for 64KB memory size, the maximum operating frequency difference between RV32I and RV32IM is the only 2MHz. For both RV32I and RV32IM implementation, the critical path is a long routing path for register bypassing from the write back stage to execute stage propagates through AGU for memory load/store. We have seen from the timing report that path delay is most significant, as BRAM blocks input/output has routed. Post place and route FPGA implementation results are shown in figure \ref{fig:rv32i_pnr}, \ref{fig:rv32im1_pnr} and \ref{fig:rv32im3_pnr} for RV32I, RV32IM(radix-4), and RV32IM(DSP) with 64KB instruction and data memory.

Compared to VexRiscv, RVCopreP-32IM(DSP) has achieved a higher frequency for both 4KB and 64KB memory. Both  VexRiscv and RVCopreP-32IM(DSP) used 4 DSP slice for the multiplier and similar logic resource utilization. And we found that RVCopreP-32IM(DSP) achieves 13\% better operating frequency than VexRiscv. 

\subsection{Benchmark}

 Dhrystone and Coremark is a synthetic benchmark program used for a long time in the industry and academy. However, Embench is relatively new as published in 2019 which is opposed to the synthetic benchmark. And they prepared a set of actual application programs to exercise branch, memory, and integer compute type instruction intensively. Embench requires a minimum of 64KB memory, that is why we used IMEM and DMEM size to 64KB for the simulation of Dhrystone, Coremark and Embench. To compile the program, riscv gcc tool version 9.2.0 is used with performance option “-O2” and machine ABI switch “-mabi=ilp32”.

\begin{table*}[htbp]

\centering

\caption{Benchmark result of RVCoreP-32IM}

\label{bench:table}

\begin{tabular}{|c|r|r|r|r|r|r|r|r|r|r|r|r|}

\hline

\multirow{3}{*}{Program name} & 

\multicolumn{4}{c|}{computation(\%)} & 

\multicolumn{2}{c|}{branch(\%)} & 

\multicolumn{2}{c|}{memory(\%)} & 

\multicolumn{2}{c|}{instr(million)} & 

\multicolumn{2}{c|}{performance gain}  \\

\cline{2-13}

    & \multirow{2}{*}{RV32I} 

    & \multicolumn{3}{c|}{RV32IM} 

    & \multirow{2}{*}{RV32I} 

    & \multirow{2}{*}{RV32IM} 

    & \multirow{2}{*}{RV32I} 

    & \multirow{2}{*}{RV32IM} 

    & \multirow{2}{*}{RV32I} 

    & \multirow{2}{*}{RV32IM}

    & \multirow{2}{*}{radix-4}  

    & \multirow{2}{*}{DSP}  \\ 

        \cline{3-5}

        &  & arith & mult  & div &  &  &  &  & & & & \\ 

\hline

dhrystone     & 44.07 & 41.39 & 0.58  & 0.30 & 21.01 & 18.49 & 34.92 & 39.24 & 0.77  & 0.68 & 0.95 & 1.02  \\

\hline

\textbf{coremark} & 58.86 & 51.37 & 3.02  & 0    & 31.48 & 22.71 &  9.67 & 22.90 & 74.48 & 31.15& 1.88 & 2.54  \\

\hline

aha-mont64    & 83.06 & 88.68 & 0.78  & 0    & 16.20 & 10.26 &  0.74 & 0.27  & 10.65 & 4.54 & 2.13 & 2.39  \\

\hline


\textbf{crc32}  & 82.31 & 69.52 & 4.34  & 0    &  8.84 & 13.05 &  8.85 & 13.08 &  5.96 & 4.03 & 0.85 & 1.48 \\

\hline

cubic         & 65.10 & 66.19 & 4.98  & 0.35 & 30.21 & 11.53 &  4.69 & 16.93 & 28.46 & 7.51 & 2.40 & 4.06  \\

\hline

edn           & 63.64 & 44.67 & 15.88 & 0.01 & 34.88 & 10.33 &  1.48 & 29.12 & 74.69 & 3.65 & 6.43 & 19.29 \\

\hline

huffbench     & 52.05 & 51.95 & 0.00  & 0.04 & 22.73 & 22.21 & 25.22 & 25.81 &  2.81 & 2.74 & 1.01 & 1.01  \\

\hline

matmult-int   & 60.71 & 38.94 & 11.06 & 0.47 & 35.24 & 12.67 &  4.06 & 36.86 & 31.15 & 3.40 & 4.62 & 9.89  \\

\hline

minver        & 61.36 & 63.74 & 1.44  & 0.68 & 31.86 & 18.03 &  6.78 & 16.09 & 13.17 & 5.08 & 2.07 & 2.46  \\

\hline

nbody         & 66.35 & 73.43 & 3.65  & 0.25 & 31.49 & 14.14 &  2.16 & 8.53  & 26.33 & 6.38 & 3.07 & 4.69  \\

\hline

nettle-aes    & 76.30 & 77.67 & 0.00  & 0.19 & 4.90  & 1.97  & 18.79 & 20.17 & 4.81  & 4.47 & 1.02 & 1.02  \\

\hline

nettle-sha256 & 84.07 & 84.08 & 0.00  & 0.00 & 1.75  & 1.73  & 14.18 & 14.19 & 4.07  & 4.07 & 0.99 & 0.99  \\

\hline

nsichneu      & 0.12  & 0.12  & 0.00  & 0.00 & 44.93 & 44.93 & 54.95 & 54.95 & 2.24  & 2.24 & 0.99 & 0.99    \\


\hline


\textbf{picojpeg}& 63.72 & 57.18 & 2.83  & 0.01 & 11.36 & 10.87 & 24.92 & 29.11 &  5.09 & 4.35 & 0.78 & 1.12    \\

\hline

qrduino       & 63.73 & 62.00 & 2.79  & 0.01 & 24.39 & 15.28 & 11.88 & 19.92 & 6.12  & 3.47 & 1.36 & 1.89    \\

\hline

sglib-combined& 38.29 & 36.50 & 0.00  & 0.39 & 27.66 & 25.70 & 34.05 & 37.41 & 2.85  & 2.60 & 0.97 & 0.97    \\

\hline

slre          & 43.30 & 43.30 & 0.00  & 0.00 & 26.23 & 26.23 & 30.47 & 30.47 & 2.74  & 2.74 & 0.99 & 0.99    \\

\hline

st            & 65.18 & 72.15 & 3.46  & 0.58 & 32.11 & 13.33 &  2.71 &  2.71 & 17.58 & 4.23 & 2.95 & 4.34    \\

\hline

statemate     & 38.72 & 38.71 & 0.00  & 0.01 & 8.67  & 8.57  & 52.60 & 52.60 & 1.75  & 1.74 & 0.99 & 0.99    \\

\hline

ud            & 58.52 & 58.05 & 3.87  & 0.90 & 28.31 & 17.99 & 13.17 & 13.17 & 5.97  & 3.33 & 1.05 & 1.46    \\

\hline

wikisort      & 57.69 & 55.18 & 2.28  & 0.94 & 31.92 & 16.83 & 10.39 & 10.39 & 6.48  & 2.57 & 1.73 & 2.21    \\

\hline

\end{tabular}

\end{table*}

 We used the Verilator tool for simulating RTL with generated program binary and measured the number of clock cycles required to execute a given benchmark program. Then execution time for the benchmark is calculated from the number of cycles required for simulation and the clock period. Clock period for RV32I, RV32IM(radix-4) and RV32IM(DSP) configuration is obtained from table \ref{fpga:table} for 64KB memory. In table \ref{bench:table}, we have presented the benchmark result. For better understanding, we have prepared the instruction histogram where we have counted the number of executed instructions by type as shown in table \ref{bench:table}.  The ratio (in percentage) of executed instructions is shown into three main categories as integer computation, branch (including jump), and memory (both load and store). For RV32IM configuration, we further subdivided the RV32I arithmetic and logical instructions into arithmetic (shown as arith in the table), multiplication instructions (shown as mult in the table), and division/reminder instructions (shown as div in the table). The total number of executed instructions (including common instructions executed during startup and end of the program) in millions is shown in the column as instr(million).

\begin{figure*}[htbp]

\begin{center}

    \includegraphics[clip,width=0.9\linewidth]{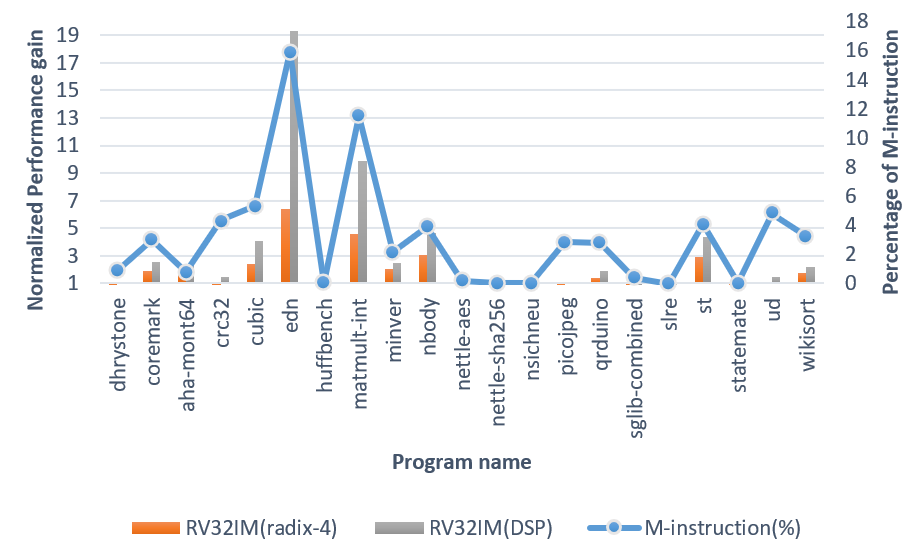}

    \caption{Achieved performance of RV32IM(radix-4) and RV32IM(DSP) compared to RV32I }

    \label{fig:gr2}

\end{center}

\end{figure*}

 Achieved performance in terms of execution time compared to RV32I is shown as a graph in figure \ref{fig:gr2}. In the same graph, the percentage of M-extension instructions that are executed for that program is shown. For any program, RV32IM(DSP) outperforms RV32IM(radix-4) because of multiplication latency.

We can see from this graph that there are several programs that do not apply or trivially apply M-extension instructions. From table \ref{bench:table}, it appears that those programs like nettle-aes, nsichneu, etc., RV32IM performance is downplayed. When there are significant numbers of M-extension instructions like cubic, edn, etc. RV32IM achieves better performance compared to RV32I. Surprisingly, programs like crc32 and picojpeg, which have used around 3\% to 4\% of M-extension instructions, but the performance gain is small compared to the program like Coremark, which executes equivalent percentage of M-extension instructions. This could be the effect of compiler optimization due to change of instruction pattern which reflects a different combination of the arithmetic, branch, and memory type instructions with and without M-extension support. However, in any of the benchmark programs, the ratio of executed divide/reminder instructions are less than 1\%.
For some programs, even though the executed number of multiplication and division instructions are few in RV32IM (for example, aha-mont64), but the total number of executed instructions is much lower than RV32I. For the given benchmark programs, we found that average performance gain RV32IM(radix-4) is 1.87, and RV32IM(DSP) is 3.13 compared to RV32I. 

We also have simulated the same benchmark program in VexRiscv and measured the execution time in milliseconds. We have compared the execution time in table \ref{bench:table2}. We found in table \ref{bench:table}, that edn, and matmult-int use the highest ratio of multiplication instructions among all of these benchmark programs. Even for those two programs, RVCoreP-32IM performs better than VexRiscv. From the table \ref{bench:table}, we can see that after accumulating of all programs execution time, VexRiscv requires 13\% more time than RVCoreP-32IM.

\begin{table}[t]

\caption{Benchmark program's execution time for RVCoreP-32IM and VexRiscv}

\label{bench:table2}

\begin{tabular}{|c|c|c|}

\hline

\multirow{2}{*}{Program name} & \multicolumn{2}{c|}{execution time (ms)}   \\ 

\cline{2-3}

                    & RVCoreP-32IM(DSP)   &  VexRiscv \\ \hline

dhrystone           &   5.33              &    6.08 \\ \hline

coremark            & 262.74              &  296.02 \\ \hline

aha-mont64          &  32.87              &   37.05 \\ \hline

crc32               &  25.94              &   35.60 \\ \hline

cubic               &  59.04              &   67.38 \\ \hline

\textbf{edn}        &  30.35              &   30.79 \\ \hline

huffbench           &  21.66              &   24.57 \\ \hline

\textbf{matmult-int}&  32.49              &   33.50 \\ \hline

minver              &  43.28              &   49.45 \\ \hline

nbody               &  48.38              &   54.74 \\ \hline

nettle-aes          &  29.23              &   33.67 \\ \hline

nettle-sha256       &  25.61              &   29.64 \\ \hline

nsichneu            &  29.12              &   29.30 \\ \hline

picojpeg            &  30.12              &   36.54 \\ \hline

qrduino             &  27.31              &   30.85 \\ \hline

sglib-combined      &  24.80              &   27.89 \\ \hline

slre                &  20.38              &   22.57 \\ \hline

st                  &  34.24              &   38.30 \\ \hline

statemate           &  11.78              &   13.62 \\ \hline

ud                  &  31.54              &   36.94 \\ \hline

wikisort            &  23.15              &   26.10 \\ \hline

Total               & 849.37              &  960.59 \\ \hline

\end{tabular}

\end{table}

We have measured the DMIPS/MHz (for dhrystone) and Coremark/MHz for three configurations of RVCoreP-32IM and the results are shown in table \ref{perf:table}.

\begin{table}[t!]

\centering

\caption{RVCoreP-32IM performance numbers}

\label{perf:table}

\centering

\begin{tabular}{|c|c|c|c|}

\hline

Benchmark    & RV32I & RV32IM(radix-4)  & RV32IM(DSP) \\

\hline

DMIPS/MHz    & 1.03  & 1.25 & 1.40   \\

\hline

Coremark/MHz & 0.84  & 1.69 & 2.33  \\

\hline

\end{tabular}

\end{table}

\section{Conclusion} \label{conclusion}


The open RISC-V ISA has opened up the opportunity to revitalize the processor microarchitecture research. As a result, the academy and industry are developing both open source and commercial core. While the RISC-V has several extensions and custom extension option, there is a key question remain how these extensions influence the computational performance of a soft processor and optimal way to integrate them. In this paper, we propose a multi-cycle execution unit of RVCoreP to support M-extension based on a simple fork-join structure in the execute stage at minimal overhead. This methodology can be used to extend the support of other RISC-V extensions like floating-point extension and so on as well as custom instructions. 

We noticed the average performance improvements of RV32IM over RV32I using the benchmark programs from Dhrystone, Coremark and Embench for in-order 5-stage pipeline architecture. Performance depends on the following factors, a) latency of multiplication and division instruction execution, b) the percentage of multiplication or division instruction executed for a given program, and c) compiler optimization on instruction with and without hardware multiplication and division. The system designer should consider the M-extension over the RV32I based on their application requirement for multiplication and division/reminder instruction. And we found that the FPGA resource utilization for RV32IM is around 1.5x of RV32I. For the given benchmark programs, the soft processor with DSP based multiplier has better performance than iterative multiplier like radix-4 booth multiplier.



\vspace{12pt}


\end{document}